\documentclass[]{jfm}

\usepackage{natbib}
\usepackage{hyperref}
\hypersetup{colorlinks=true,urlcolor=blue,citecolor=black}
\usepackage{pgfplots}
\usetikzlibrary{shapes, arrows, calc, positioning}
\usepackage{graphicx}
\usepackage{amsmath}
\usepackage{soul} 
\usepackage{tikz}
\usepackage{longtable}
\usepackage{siunitx}

\usepackage{newtxtext}
\usepackage{newtxmath}

\graphicspath{{figs/}} 

\definecolor{re1000}{RGB}{0,255,0}  
\definecolor{re2000}{RGB}{0,0,255}
\definecolor{re3000}{RGB}{0,255,255}
\definecolor{re6000}{RGB}{255,0,0}

\newcommand{\aver}[1]{\langle #1 \rangle}
 
\newcommand{\DR}{\mathcal{R}}
\newcommand{\NDR}{\mathcal{S}}
\newcommand{\cfa}{C_f}
\newcommand{\cfr}{C_{f_{\; 0}}}
\newcommand{\pin}{P_{\mathrm{in}}}
\newcommand{\ppnot}{P_{p_{0}}}
\newcommand{\dd}{\mathrm{d}}

\title[]{Turbulent skin-friction drag reduction\\ via spanwise forcing at high Reynolds number}

\author{Davide Gatti
\aff{1} \corresp{\email{davide.gatti@kit.edu}},
Maurizio Quadrio
\aff{2},
Alessandro Chiarini
\aff{2,3},
Federica Gattere
\aff{2}
and 
Sergio Pirozzoli
\aff{4}
}
\affiliation{
\aff{1} Institute of Fluid Mechanics, Karlsruhe Institute of Technology,
    Kaiserstra\ss e 10, 76131 Karlsruhe, Germany \\
\aff{2} Dipartimento di Scienze e Tecnologie Aerospaziali, Politecnico di Milano,
via La Masa 34, 20156 Milano, Italy \\
\aff{3} Complex Fluids and Flows Unit, Okinawa Institute of Science and Technology Graduate University,
1919-1 Tancha, Onna-son, Okinawa 904-0495, Japan \\
\aff{4} Dipartimento di Meccanica e Aeronautica, Universita di Roma ``La Sapienza'', `
Via Eudossiana 18, 00184 Rome, Italy \\
}

\begin{document}
\maketitle

\begin{abstract}

We address the Reynolds-number dependence of the turbulent skin-friction drag reduction induced by streamwise-travelling waves of spanwise wall oscillations. 
The study relies on direct numerical simulations of drag-reduced flows in a plane open channel at friction Reynolds numbers in the range $1000 \le Re_\tau \le 6000$, which is the widest range considered so far in simulations with spanwise forcing.
Our results corroborate the validity of the predictive model proposed by \cite{gatti-quadrio-2016}: regardless of the control parameters, the drag reduction decreases monotonically with $\Rey$, at a rate that depends on the drag reduction itself and on the skin-friction of the uncontrolled flow. 
We do not find evidence in support of the results of \cite{marusic-etal-2021}, which instead report by experiments an increase of the drag reduction with $\Rey$ in turbulent boundary layers, for control parameters that target low-frequency, outer-scaled motions. 
Possible explanations for this discrepancy are provided, including obvious differences between open channel flows and boundary layers, and possible limitations of laboratory experiments.

\end{abstract}

\section{Introduction}
\label{sec:introduction}

Transverse near-wall forcing as a means to mitigate skin-friction drag in turbulent flows has gathered significant attention, owing to its potential for substantial environmental and economic benefits \citep{quadrio-2011, ricco-skote-leschziner-2021}. 
After the seminal work on spanwise wall oscillations by \cite{jung-mangiavacchi-akhavan-1992}, three decades of research efforts have led to important progress; however, several crucial factors still hinder the deployment of spanwise forcing in technological settings.
The major challenge resides in devising viable and efficient implementations of the typically idealised near-wall forcing, but other concerns exist, including the decreasing effectiveness of drag reduction with increasing Reynolds numbers ($\Rey$). 

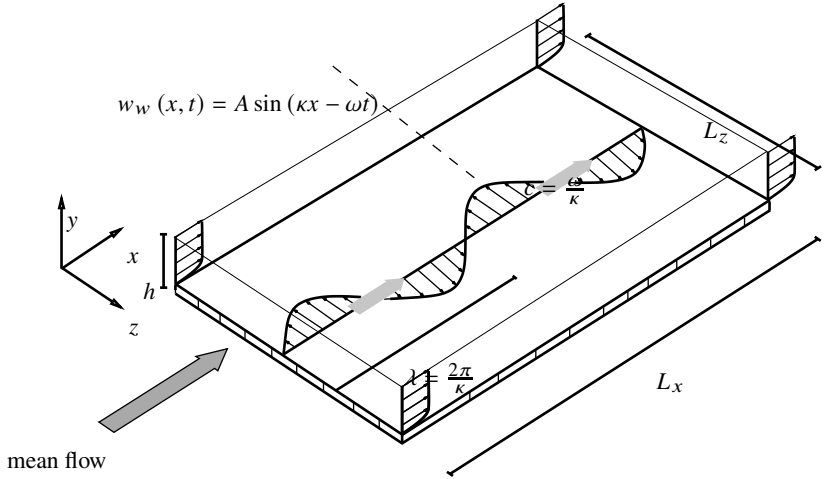
\begin{figure}
\begin{tikzpicture}
    \node[anchor=south west,inner sep=0] (image) at (0,0) {\includegraphics[width=0.75\textwidth]{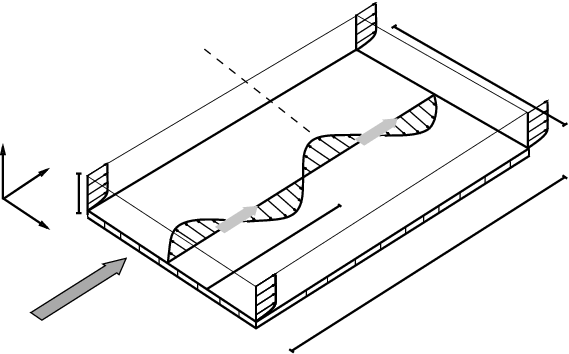}};
    \begin{scope}[x={(image.south east)},y={(image.north west)}]
        \node[] at (0.10,0.46) {$x$};
        \node[] at (0.02,0.53) {$y$};
        \node[] at (0.10,0.31) {$z$};
        \node[] at (0,0.04) {mean flow};
        \node[] at (0.8,0.20) {$L_x$};
        \node[] at (0.86,0.72)   {$L_z$};
        \node[] at (0.12,0.39) {$h$};
        \node[] at (0.5,0.21) {$\lambda =\frac{2\pi}{\kappa}$};
        \node[] at (0.65,0.6) {$c=\frac{\omega}{\kappa}$};
        \node[] at (0.25,0.78) {$w_w\left(x,t\right)=A\sin \left(\kappa x - \omega t \right)$};
    \end{scope} 
\end{tikzpicture}
 \caption{Schematic of a turbulent open channel flow actuated with streamwise-travelling waves of spanwise wall velocity with amplitude $A$, streamwise wavenumber $\kappa$ and angular frequency $\omega$. Here, $\lambda$ is the streamwise wavelength; $c$ is the wave phase speed; and $L_x$, $L_y=h$ and $L_z$ are the dimensions of the computational domain in the streamwise, wall-normal and spanwise direction, respectively. }
\label{fig:travellingWaves}
\end{figure}

To date, the Reynolds dependence of skin-friction drag reduction has mostly been studied in the context of streamwise-travelling waves of spanwise wall velocity \citep[StTW,][]{quadrio-ricco-viotti-2009}, a specific form of transverse forcing characterised by its comparatively large potential for drag reduction with modest energy expenditure. 
StTW are described by 
\begin{equation}
w_w (x,t) = A \sin \left( \kappa x - \omega t \right),
\label{eq:travellingWaves}
\end{equation}
where $w_w$ is the spanwise ($z$) velocity component at the wall, $A$ is the maximum wall velocity and thus a measure of the amplitude of the spanwise forcing,  $\kappa$ is the streamwise wavenumber, $\omega$ is the angular frequency, and $x$ and $t$ are the streamwise coordinate and the time. 
The forcing, sketched in figure~\ref{fig:travellingWaves}, consists of streamwise-modulated waves of spanwise velocity at the wall, with wavelength $\lambda = 2\pi / \kappa$ and period $T = 2\pi / \omega$. The waves travel along the streamwise direction with phase speed $c=\omega/\kappa$, either downstream ($c>0$) or upstream ($c<0$) with respect to the mean flow direction.
The forcing described by equation~\eqref{eq:travellingWaves} includes the two special cases of spatially uniform spanwise wall oscillations \citep{quadrio-ricco-2004} for $\kappa=0$, and steady waves \citep{viotti-quadrio-luchini-2009} for $\omega=0$. 
With the appropriate set of control parameters, StTW have been shown to yield considerable drag reduction in a series of numerical experiments regarding channel and pipe flows \citep{quadrio-ricco-viotti-2009, gatti-quadrio-2013, hurst-yang-chung-2014, gatti-quadrio-2016, liu-etal-2022, rouhi-etal-2023, gallorini-quadrio-2024} and boundary layers \citep{skote-schlatter-wu-2015, skote-2022}, as well as in laboratory experiments \citep{auteri-etal-2010, bird-santer-morrison-2018, chandran-etal-2023}. Besides canonical flows, including the compressible and supersonic regimes \citep{gattere-etal-2024}, StTW have been applied to more complex flows ranging from channels with curved walls \citep{banchetti-luchini-quadrio-2020}, to rough boundary layers \citep[][although restricted to spatially uniform spanwise wall oscillation]{deshpande-etal-2023b} and transonic airfoils with shock waves \citep{quadrio-etal-2022}, showing that local skin-friction drag reduction can be exploited to also reduce the pressure component of the aerodynamic drag.

Understanding how the Reynolds number affects drag reduction by StTW is a particularly challenging goal for three main reasons. First, a sufficiently wide portion of a huge parameter space must be explored, which even in simple canonical flows includes the four parameters $\left\{ A, \kappa, \omega; Re \right\}$, and poses a great challenge to numerical and laboratory experiments. 

A second complication is the choice of an appropriate figure of merit for drag reduction. Typically, the drag reduction rate $\DR$ is defined as
\begin{equation}
    \DR = 1 - \frac{\cfa}{\cfr} \, ,
    \label{eq:dr}
\end{equation}
i.e. as the control-induced relative change of the skin-friction coefficient $\cfa$ \citep{kasagi-hasegawa-fukagata-2009}. In equation~\eqref{eq:dr} and in the remainder of this manuscript, the subscript $0$ denotes quantities measured in the reference uncontrolled flow. Specifically, $\cfa$ is defined as $\cfa = 2 \tau_x / (\rho U_b^2$); $\tau_x$ is the mean streamwise wall shear stress, $U_b$ the bulk velocity, and $\rho$ the fluid density. 
However, as observed by \cite{gatti-quadrio-2016}, the quantity $\DR$ defined by equation~\eqref{eq:dr} is inherently $\Rey$-dependent, owing to the $\Rey$-dependence of $\cfa$ and $\cfr$. This is long known to be the case for the flow over rough surfaces \citep{nikuradse-1933, jimenez-2004}, as well as for other flow control techniques relying on near-wall turbulence manipulation such as riblets \citep{luchini-1996, spalart-mclean-2011}. 
Choosing a figure of merit which eliminates this trivial dependency on the Reynolds number is crucial to describe properly the $\Rey$-effect on drag reduction.

Third, the wall shear stress generally differs in the reference ($\tau_{x_0}$) and controlled ($\tau_x$) channel flows, unless they are driven by the same pressure gradient \citep[as done for example by][]{ricco-etal-2012}; the viscous scaling, hence, becomes ambiguous. 
As noted by \cite{quadrio-2011}, this results in two possible viscous normalisations of the controlled flow: the first, denoted with the superscript `$+$', relies on the reference friction velocity $u_{\tau_0} = \sqrt{\tau_{x_0} / \rho}$; the second, denoted with the superscript `$\ast$', is based on the actual friction velocity $u_\tau =  \sqrt{\tau_x / \rho}$. 
Similarly, two different friction Reynolds numbers, $\Rey_{\tau_0}={u_{\tau_0} h} / \nu$ and $\Rey_\tau={u_\tau h} / \nu$ can be defined depending on the choice of the friction velocity. Here, $h$ describes the half-height of a channel or the depth of an open channel, and $\nu$ is the fluid kinematic viscosity. 
While the actual viscous scaling is the only sensible choice for the drag-reduced flow \citep{gatti-quadrio-2016}, the reference scaling is necessary when the wall friction of the drag-reduced flow is not known yet. 

\cite{gatti-quadrio-2016}, indicated also as GQ16 hereinafter, circumvented these difficulties by designing a campaign of several thousands direct numerical simulations (DNS) of turbulent channel flows. 
Inspired by similar studies on rough walls \citep[see for example][]{leonardi-etal-2015}, they limited the otherwise prohibitive computational cost by choosing relatively small computational domains \citep{jimenez-moin-1991,flores-jimenez-2010} for most of the study.
At the expense of a residual domain-size dependence of the results, which cancels out in large part when observing the difference between controlled and uncontrolled flows, GQ16 generated a large dataset, 
along with a more limited number of simulations in wider domains to verify the accuracy of the results.
This approach enabled not only the inspection of a large portion of the $\left\{ A, \kappa, \omega \right\}$-space at $\Rey_{\tau_0} = 200$ and 1000, but also the transfer of the dataset between viscous `$+$' and `$\ast$' units via interpolation, allowing to assess the results in both scalings. 
Thanks to their comprehensive database (available as Supplementary Material to their paper), \cite{gatti-quadrio-2016} challenged the then-current belief that skin-friction drag reduction was bound to decrease quickly with $\Rey$. They demonstrated that the drag reduction effect by spanwise forcing becomes in fact constant with $\Rey$, provided that it is not expressed via $\DR$ (equation~\eqref{eq:dr}), that is {\em per se} $\Rey$-dependent, but through the Reynolds number-invariant parameter $\Delta B^*$. The quantity $\Delta B^*$ expresses the main effect of the StTW, which is to induce a change of the additive constant in the logarithmic law for the mean velocity profile
\begin{equation}
    U^*(y^*) = \frac{1}{k} \ln y^* + B_0^* + \Delta B^* \, ,
    \label{eq:loglaw}
\end{equation}
where $k$ is the von Kármán constant, $B_0^*$ is the additive constant in the reference channel flow, and $B^*=B_0^*+\Delta B^*$ is the additive constant of the controlled flow. 
The independency of $\Delta B^*$ upon $\Rey$ is a common feature of all turbulence manipulations whose action is confined to the near-wall region. In these cases the outer turbulence simply reacts to a wall layer with different drag \citep{gatti-etal-2018b}, as well known, for instance, in the context of drag-reducing riblets \citep{luchini-1996, garcia-jimenez-2011, spalart-mclean-2011} and drag-increasing roughness \citep{clauser-1954, hama-1954}. 

Under the assumption that $\Delta B^*$ is a function of the control parameters $\left\{ A^\ast, \kappa^\ast, \omega^\ast \right\}$, but not of the Reynolds number, \cite{gatti-quadrio-2016} derived the following modified friction relation (hereinafter called GQ model)    
\begin{equation}
    \Delta B^* = \sqrt{\frac{2}{\cfr}} \left[ (1-\DR)^{-1/2} -1 \right] - \frac{1}{2 k}\ln \left(1 - \DR \right) \, ,
    \label{eq:gqmodel}
\end{equation}
where the $\Rey$-dependence is not explicit, but rather embedded in $\cfr$. 
Provided the function $\Delta B^* \left(A^\ast, \kappa^\ast, \omega^\ast \right)$ is measured at a sufficiently large $\Rey$ for the log law in equation~\eqref{eq:loglaw} to hold, the GQ model predicts $\DR$ at any arbitrary value of $\Rey$. 
According to equation~\eqref{eq:gqmodel}, $\DR$ is always expected to decrease with $\Rey$ for any combination of the control parameters, but at much lower rate than suggested by previous studies \citep{touber-leschziner-2012,gatti-quadrio-2013,hurst-yang-chung-2014}, so that significant drag reduction can be still achieved at Reynolds numbers typical of technological applications. For instance, for StTW GQ16 estimated possible drag reduction of 30\% with $A^+=12$ at $\Rey_{\tau_0}=10^5$. 

The GQ16 study is affected by two limitations. 
First, $\Rey_{\tau_0}=1000$, the largest $\Rey$ considered in their study, may still be not enough for $\Delta B^*$ to become completely $\Rey$-independent: GQ16 suggested that at least $\Rey_{\tau_0}=2000$ should be considered. Second, the small effect of the restricted computational box sizes on the quantification of $\DR$ could in principle bias the extrapolation to higher $\Rey$. 
Nonetheless, the GQ model passed validation tests against previous \citep{touber-leschziner-2012, hurst-yang-chung-2014} and later literature data. For instance, \cite{rouhi-etal-2023} employed large eddy simulations (LES) to study drag reduction by StTW in open channel flows at $\Rey_{\tau_0}=945$ and $\Rey_{\tau_0}=4000$. They explored the parameter space within the range $\kappa^+ \in \left[0.002, 0.02 \right] $ and $\omega^+ \in \left[-0.2, 0.2\right] $, at fixed $A^+=12$. This is to be compared with $\kappa^+ \in [0,0.05]$ and $\omega^+ \in [-0.5, 1]$ addressed by \cite{gatti-quadrio-2016}, who also considered various amplitudes $A^+ \in [2, 20]$. 
The study of \cite{rouhi-etal-2023} is however limited by the use of large eddy simulations (LES), in which part of the small-scale turbulence physics involved in drag reduction is modelled, and by the domain size ($L_x = 2.04h$,  $L_z = 0.63h$ at $\Rey_{\tau_0}=4000$), which is comparable to the restricted domain size ($L_x = 1.35h$,  $L_z = 0.69h$ at $\Rey_{\tau_0}=1000$) considered by \cite{gatti-quadrio-2016}, despite the larger $\Rey_{\tau_0}$. 
\cite{rouhi-etal-2023} confirmed that the GQ model predicts very well their drag reduction data, with deviations in the order of 2\%, for all StTW control parameters sufficiently far from those yielding drag increase. 

\cite{marusic-etal-2021} and \cite{chandran-etal-2023} studied drag reduction via backward-travelling ($c<0$) StTW. Their experimental study was carried out in a zero pressure gradient turbulent boundary layer up to the largest values of $\Rey$ investigated so far, $\Rey_\tau = 15000$. 
By extending to the plane geometry the actuation strategy used by \cite{auteri-etal-2010} in a cylindrical pipe, they implemented the ideal forcing of equation~\eqref{eq:travellingWaves} by dividing a portion of the wall into a series of forty-eight slats, long $5\, \mathrm{cm}$ each, so that each six consecutive slats constitute a single wavelength with fixed $\lambda = 0.3\, \mathrm{m}$. The slats move in the spanwise direction at a fixed half-stroke $d$, resulting in a frequency-dependent maximum spanwise velocity $A = \omega d$. As a consequence in those experiments the amplitude and period of the oscillations could not be varied independently. With $d$ and $\lambda$ constant in physical units, the range of investigated parameters shifts towards smaller $\kappa^+$, $\omega^+$ and $A^+$ as $\Rey_{\tau_0}$ increases. 
The authors observed, for the first time, $\DR$ to increase with $\Rey$ \citep[see figure 3{\it e} of][]{marusic-etal-2021}, and explained it with the particularly slow timescale $T^+ = 2 \pi / \omega^+ < -350$ of their forcing, which was meant to target the large inertial, outer-scaled structures of turbulence \citep{deshpande-etal-2023c}, whose importance increases with $\Rey$. 

Despite the promising results, these studies also have shortcomings. With $d$ and $\lambda$ constant in physical units, which is unavoidable in laboratory experiments, the control parameters could not be kept constant in either `$+$' or `$*$' viscous units while varying $\Rey_{\tau_0}$. In particular, the fixed wavelength leads to a $\kappa^+$ that decreases with $\Rey$.
Furthermore, the effect of $\omega$ and $A$ cannot be addressed separately. This precludes the investigation of the full space of the control parameters: for instance large values of $\omega^+$ at low $A^+$ cannot be tested. Lastly, the key observation that $\DR$ increases with $\Rey$ relies on the joint observation of low-$\Rey$ LES data by \cite{rouhi-etal-2023} obtained in an open channel flow, and high-$\Rey$ experimental data by \cite{marusic-etal-2021} in a boundary layer, thus bringing together different methods and flow configurations.

The present research fills these gaps in the existing literature by leveraging a novel DNS dataset of turbulent open channel flow, to accurately quantify the Reynolds number effects on the drag-reducing performance of StTW.  
The computational domain adopted in the present simulations is large enough to properly account for all relevant scales of turbulence, including the large inertial scales. The considered Reynolds numbers, ranging from $\Rey_{\tau_0}=1000$ to $\Rey_{\tau_0}=6000$, are large enough to minimise the low-$\Rey$ effects, matching some of the experimental data points by \cite{chandran-etal-2023}. The dataset is further designed to address the Reynolds-number scaling of drag reduction in both viscous and outer units independently, by considering the same flow configuration and by using the same numerical method for all $\Rey$.

The paper is organised as follows. After this Introduction, \S\ref{sec:methods} describes the computational procedure and the simulation parameters used to produce the DNS dataset. In \S\ref{sec:res1} the effect of the Reynolds number is analysed in terms of both drag reduction and power budgets, and compared to existing literature. Finally, concluding arguments are given in \S\ref{sec:conclusion}.
 \section{Methods and procedures}
\label{sec:methods}

\begin{table}
    \begin{tabular*}{\textwidth}{@{\extracolsep{\fill} } l c c c c c c }
    $\Rey_b$ & $\Rey_{\tau_0}$ & $N_\mathrm{cases}$ & $L_x / h$ & $L_z / h$ &  $N_x \times N_y \times N_z$      & Symbol \\
    \hline
    20000  &         996.7 &   71    & 6$\pi  h$ & 2$\pi  h$ &  2304 $\times$ 165 $\times$ 1536  & {\color{re1000} $\blacktriangle$} \\
    43650  &        1994.1 &   62     & 6$\pi  h$ & 2$\pi  h$ &  4608 $\times$ 265 $\times$ 3072  & {\color{re2000} $\blacktriangledown$} \\
    68600  &        3008.8 &   7     & 6$\pi  h$ & 2$\pi  h$ &  6912 $\times$ 355 $\times$ 4608  & {\color{re3000} $\blacklozenge$} \\
    148000  &       6012.6 &   3     & 6$\pi  h$ & 2$\pi  h$ &  13312 $\times$ 591 $\times$ 9216  & {\color{re6000} $\Large\bullet$} \\
    \end{tabular*}
\caption{Details of the direct numerical simulations of open channel flows (including domain size and discretization) modified by StTW, grouped in sets of $N_\mathrm{cases}$ simulations performed at a constant value of bulk Reynolds number $\Rey_b = U_b h / \nu$. 
The last column indicates the color and symbol employed in the following figures to represent each set of simulations.}
    \label{tab:parameters}
\end{table}

A new DNS dataset of incompressible turbulent open-channel flows (see figure~\ref{fig:travellingWaves}) is used to study the effect of the Reynolds number on the reduction of the turbulent friction drag achieved by StTW. 
The open channel flow, i.e. half a channel flow with a symmetry boundary condition at the centreplane, is considered here to reduce the computational cost without affecting the drag reduction results; indeed, it was often used in the past, including e.g. the similar studies by \cite{yao-chen-hussain-2022}, \cite{pirozzoli-2023} and \cite{rouhi-etal-2023}. 
The StTW are applied as a wall boundary condition for the spanwise velocity component after equation~\eqref{eq:travellingWaves}. Periodic boundary conditions are applied in the homogeneous streamwise and spanwise directions, no-slip and no-penetration boundary conditions are used for the longitudinal and wall-normal components at the bottom wall; free slip is used at the top boundary. 
The computational setup is identical to the study of \citet{pirozzoli-2023}, in which open-channel flow was studied in the absence of flow control. The solver relies on the classical fractional step method with second-order finite differences on a staggered grid \citep{orlandi-2006}. The Poisson equation resulting from the divergence-free condition is efficiently solved via Fourier expansion in the periodic directions \citep{kim-moin-1985}. The governing equations are advanced in time starting from the initial condition of a statistically stationary, uncontrolled turbulent open channel flow by means of a hybrid third-order, low-storage Runge--Kutta algorithm, whereby the diffusive terms are handled implicitly. Statistical averaging, indicated hereinafter as $\aver{ \cdot }$, implies averaging in time and along the two homogeneous directions.

Four sets of simulations, whose details are listed in table \ref{tab:parameters}, are run at prescribed values of the bulk Reynolds number $\Rey_b = U_b h / \nu$; the bulk velocity is kept constant at every time step as described in \cite{quadrio-frohnapfel-hasegawa-2016}. Each set comprises one reference simulation, in which the wall is steady, and a variable number of cases with StTW  at different values of $\left\{A, \kappa, \omega \right\}$. 
In the following, we will refer to each simulation set via its (nominal) value of $\Rey_{\tau_0}$; the actual values of $\Rey_\tau$ vary throughout simulations of each set, as a consequence of the wall actuation at constant $U_b$. 
 
All DNS are carried out in a domain with $L_x = 6 \pi h$ and $L_z = 2 \pi h$, which is much larger than what has been adopted by \cite{rouhi-etal-2023} and GQ16 at similar values of $\Rey$, but a bit smaller than the domain used by \cite{yao-chen-hussain-2022}. Whereas weak longitudinal eddies may be not resolved, a box sensitivity study carried out by \citet{pirozzoli-2023} showed
that the practical impact on the leading-order flow statistics and on the spanwise spectra is extremely small.

\begin{figure}
  \centering
  \hspace{-10pt}
  \includegraphics[]{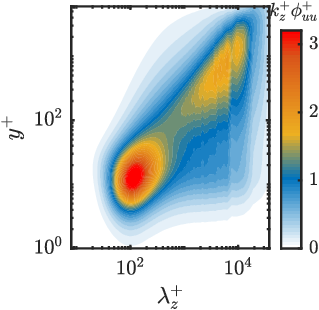}
  \hspace{30pt}
  \includegraphics[]{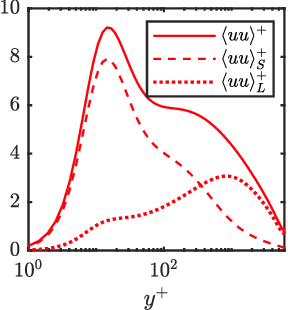}
  \caption{Statistics of streamwise velocity fluctuations for the reference simulation at $\Rey_{\tau_0} = 6000$: (left) spanwise premultiplied spectra $k_z^+ \phi_{uu}^+$; (right) streamwise variance $\aver{uu}^+$ with its large-scale $\aver{uu}_L^+$ and small-scale $\aver{uu}_S^+$ contributions. Large scales are defined as those for which $2 \pi / k_z > 0.5h$.}
\label{fig:spectra}
\end{figure}
Figure \ref{fig:spectra} indeed supports the adequacy of the present computational box by analysing the streamwise velocity fluctuations of the reference open channel flow at $\Rey_{\tau_0}=6000$, i.e. the largest Reynolds number considered in the present study. Figure \ref{fig:spectra} (left) shows the spanwise pre-multiplied spectrum $k_z^+ \phi_{uu}^+$, where $k_z$ is the spanwise wavenumber and $\phi_{uu}$ is a component of the velocity spectrum tensor, with a clear outer peak visible at $\lambda_z \approx h$. Figure \ref {fig:spectra} (right) shows the variance $\aver{uu}^+$ of the streamwise velocity, split into the large-scale $\aver{uu}^+_L$ and small-scale $\aver{uu}^+_S$ contributions. The large-scale contribution is obtained by integrating the spectrum only for wavelengths $\lambda_z > 0.5h$ as suggested by \cite{bernardini-pirozzoli-2011}, \cite{dogan-etal-2019} and \cite{yao-chen-hussain-2022}. 
With this definition, the large-scale fluctuations are responsible for 12\% of the total variance in the vicinity of the wall, and for as much as 85\% at the free-slip surface.  Moreover, it should be noted that the longest travelling wave that we have tested at the highest Reynolds number ($\Rey_{\tau_0}=6000$) is fourteen times shorter than the domain length, thus allowing subharmonic effects, if present, to be properly resolved. 

The spatial resolution of the simulations is designed based on the criteria discussed by \cite{pirozzoli-orlandi-2021}. In particular, the collocation points are distributed in the wall-normal direction $y$ so that approximately thirty points are placed within $y^+ \leq 40$, with the first grid point at $y^+ < 0.1$. 
The mesh is stretched in the outer wall layer with the mesh spacing proportional to the local Kolmogorov length scale, which there varies as $\eta^+ \approx 0.8 (y^+)^{1/4}$ \citep{jimenez-2018}. A mild refinement towards the free surface is used in order to resolve the thin layer in which the top boundary condition dampens the wall-normal velocity fluctuations. The grid resolution in the wall-parallel directions is set to $\Delta x^+ \approx 8.5$ and $\Delta z^+ \approx 4.0$ for all the flow cases. Note that the resolution is finer in actual viscous units in all cases with drag reduction. 

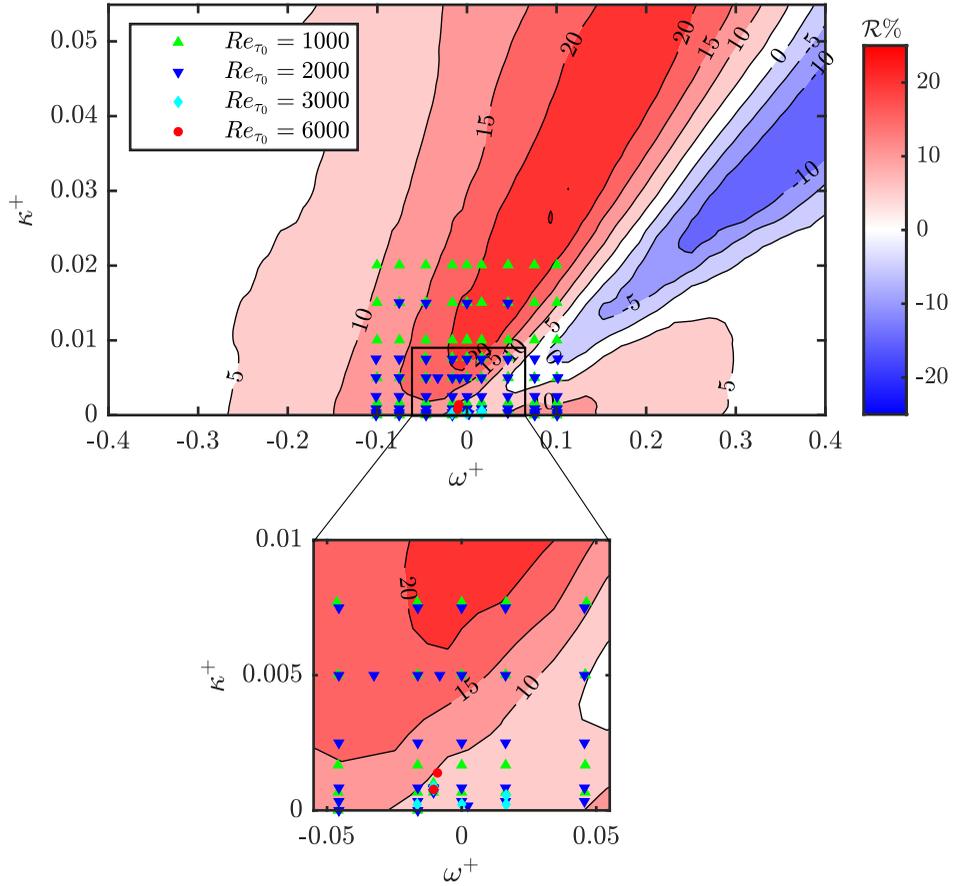
\begin{figure}
  \begin{tikzpicture}
    \begin{scope}[xshift=1.5cm]
    \node[anchor=south west,inner sep=0] (image) at (0,0) {\includegraphics[]{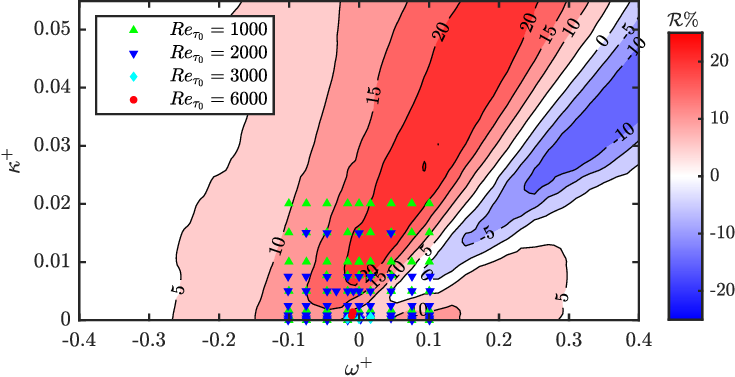}};
    \node[anchor=north,inner sep=0, below=of image, xshift=-9mm, yshift=5mm] (image2) {\includegraphics[]{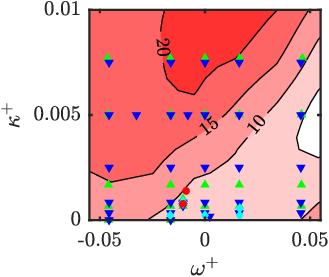}};
    \begin{scope}[x={(image.south east)},y={(image.north west)}]
      \draw[black, thick] (0.43,0.15) rectangle (0.55,0.29);
      \draw [] (0.43,0.15) -- (0.325,-0.11);
      \draw [] (0.55,0.15) -- (0.64,-0.11);
    \end{scope}
  \end{scope}
\end{tikzpicture}\caption{Portion of the parameter space spanned in the present study overlaid to the drag reduction map by GQ16 computed at $A^+=5$. Each symbol corresponds to one simulation at the Reynolds number encoded by its shape/color, as described by the legend.}
\label{fig:parameterSpace}
\end{figure}
Figure \ref{fig:parameterSpace} shows at a glance the range of the StTW parameters addressed in the present study for the simulation sets at $\Rey_{\tau_0}= \left\{1000, 2000, 3000, 6000 \right\}$. This is the broadest range of $\Rey$ considered so far in numerical simulations with spanwise wall forcing. 

The portion of the $\left\{\kappa^+, \omega^+ \right\}$-space spanned in the present study is smaller than the one addressed in GQ16. In fact,  we limit ourselves to considering $\kappa^+ \leq 0.02$ and $\left| \omega^+ \right| \leq 0.1$, which is now known to be the most interesting part of the parameter space, where the maxima of drag reduction $\DR$ and net saving $\NDR$ are expected.  

The control parameters have been selected according to the following guiding principles. 
\begin{itemize}
    \item[(1)] The intent to further scrutinize the validity of the results by GQ16, obtained in constrained computational domains, led us to consider a wider portion of the StTW parameter space at $\Rey_{\tau_0}=1000$, the highest value considered in their study. 
    
    \item[(2)] GQ16 observed that $\Delta B^*$ may still retain residual dependence on $\Rey$ at their highest value of $\Rey_{\tau_0}=1000$, and suggested that at least $\Rey_{\tau_0}=2000$ is needed for a $\Rey$-independent measure. Therefore, the same region of the parameter space considered above in point (1) is also considered at $\Rey_{\tau_0}=2000$.
    
    \item[(3)] \label{item:iii} \cite{marusic-etal-2021} reported for the first time a drag reduction that increases with $\Rey$ for small values of $\kappa^+$ and $\omega^+$, in particular for $\kappa^+=0.0008$ (i.e. $\lambda^+ \approx 8000$), $\omega^+=-0.0105$ (i.e. $T^+ \approx -600$) and $A^+ \approx 5$ (in fact their $A^+$ varies slightly across the $\Rey$-number range), as shown in figure 3{\it e} of their paper. We have added this combination of $\left\{ \kappa^+, \omega^+ \right\}$ to all simulations sets, in order to verify the increase of $\DR$ with $\Rey$. This is one of the two controlled cases we have carried out at $\Rey_{\tau_0}=6000$. The second case, with $\kappa^+=0.0014$, $\omega^+=-0.009$ and $A^+=2.5$, matches exactly one of the cases considered experimentally by \cite{chandran-etal-2023}, at the same value of $\Rey_{\tau_0}=6000$.
 
    \item[(4)]  All controlled simulations are performed at $A^+=5$ for two reasons: first, this value of $A^+$ is representative of the amplitude range in the experiments by \cite{marusic-etal-2021} for the case discussed at point (3); second, this value is close to $A^+ \approx 6$ at which GQ16 measured the maximum of net power saving $\NDR$. By adopting this value of $A^+$ we can verify whether positive $\NDR$ can also be achieved at higher $\Rey$. 
\end{itemize}

This results in the combination of the control parameters shown in figure \ref{fig:parameterSpace}, and listed in tables from \ref{tab:ret1000}, \ref{tab:ret2000}, \ref{tab:ret3000} and \ref{tab:ret6000} of appendix \ref{sec:appendix} together with the main results.
As will be clarified in the following, understanding the $\Rey$-dependence of $\DR$ and $\NDR$ requires accurate estimation of the mean wall friction, which we guarantee by monitoring statistical uncertainty via the method described by \cite{russo-luchini-2017}, as shown in figures \ref{fig:dr_re} and \ref{fig:s_re}. Statistics are accumulated for at least $10 h / u_{\tau_0}$ time units after the initial transient, during which the control leads the flow towards a reduced level of drag.
 \section{Results}
\label{sec:res1}

The outcomes of the present study are presented following the guiding principles outlined in \S\ref{sec:methods}. First, we present drag reduction maps at $\Rey_{\tau_0}=1000$ and 2000 and use them to provide ultimate validation of the GQ16 results. Second, we evaluate $\Delta B^\ast$ at $\Rey_{\tau_0} = 2000$ and verify the $\Rey$-independence of this drag reduction metric. Third, drag reduction is reported up to $\Rey_{\tau_0}=6000$ for the same actuation parameters for which \cite{marusic-etal-2021} observed drag reduction increase with $\Rey$. 
Finally, the possibility to achieve net power savings at high $\Rey$ is discussed. 

\subsection{Maps of $\DR$: validity of the results by GQ16}
\begin{figure}
    \centering
    \includegraphics[]{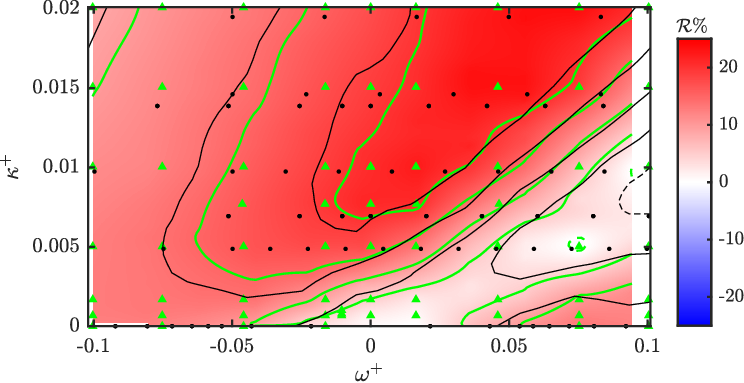}
    \includegraphics[]{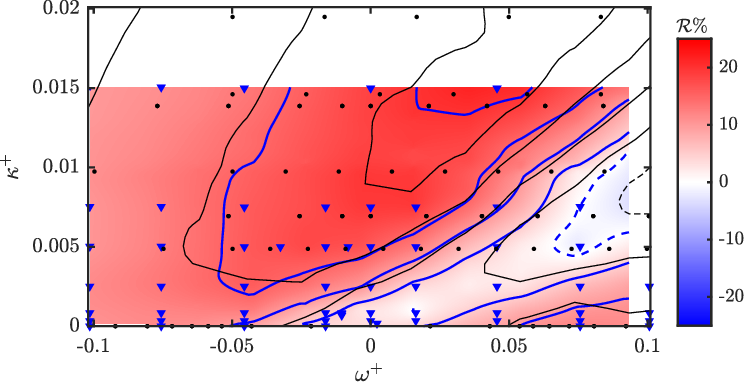}
\caption{Maps of drag reduction ($\DR$) as a function of actuation parameters ($\omega^+$, $\kappa^+$), at $\Rey_{\tau_0}=1000$ (top) and $\Rey_{\tau_0}=2000$ (bottom). The colormap, the contour lines and symbols colored after table \ref{tab:parameters} refer to the present data, whereas the black contour lines and symbols refer to the data by GQ16, which at $\Rey_{\tau_0}=2000$ are obtained from extrapolation through GQ model \eqref{eq:gqmodel}. The contour lines are every 5\% of $\DR$, dashed lines mark the $\DR=0$ iso-line. }
    \label{fig:dr}
\end{figure}

Figure~\ref{fig:dr} compares the present drag reduction results at $\Rey_{\tau_0}=1000$ and $\Rey_{\tau_0}=2000$  with the data by GQ16, which need to be transferred to the present values of $\Rey_{\tau_0}$. The procedure involves starting from their $\DR$ and $\cfr$ data, then using the GQ model (equation \ref{eq:gqmodel} with $k=0.39$; GQ16 showed that the specific value of $k$ in the range $0.385$--$0.4$ does not significantly affect the results) to compute $\Delta B^\ast$. The resulting cloud of $\Delta B^\ast$ data points at discrete $\left\{ A^+, \kappa^+, \omega^+ \right\}$ values is linearly interpolated on a Cartesian grid spanning the $\left\{ \kappa^+, \omega^+ \right\}$ space at the value of $A^+=5$ considered in the present study. Finally, $\Delta B^\ast$ is again converted back to $\DR$ values via the GQ model, now with the values of $\cfr$ corresponding to $\Rey_{\tau_0}=1000$ and $\Rey_{\tau_0}=2000$. 

The comparison shows excellent agreement between the two datasets. 
This finding suggests very weak sensitivity of StTW actuation on the flow geometry (open channel vs. closed plane channel), and further strengthens the reliability of the GQ16 data. In fact, due to their limited domain size, GQ16 had no data for $0 < \kappa^+ < 0.005$, but even there the new data compare very well with the GQ16 map. 
The maximum difference between the present and GQ16 datasets evaluated across the interpolated maps shown in figure \ref{fig:dr} is only 2.5\%, and the standard deviation is 0.8\%. 
The agreement shows that no measurable direct effect of large-scale turbulent structures on $\DR$ exists at these values of $\Rey_{\tau_0}$ other than their possible contribution to $\cfr$, which is already accounted for by the GQ model. 

\subsection{Maps of $\Delta B^\ast$: validity of the GQ model}

\begin{figure}
    \centering
    \includegraphics[]{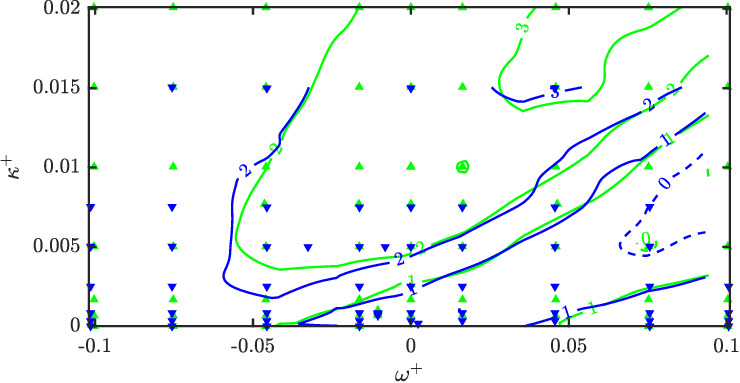}
	\caption{Maps of $\Delta B^\ast$ as a function of actuation parameters ($\omega^+$, $\kappa^+$) at $\Rey_{\tau_0}=1000$ \mbox{(
        \protect\tikz{\protect\draw[thick,re1000] (0pt,2pt) -- (20pt,2pt); 
                      \protect\draw[thick,white] (0pt,0pt) -- (20pt,0pt);}
        )} and $\Rey_{\tau_0}=2000$ \mbox{(
        \protect\tikz{\protect\draw[thick,re2000] (0pt,2pt) -- (20pt,2pt); 
                      \protect\draw[thick,white] (0pt,0pt) -- (20pt,0pt);}  
        )}. The symbols are colored after table \ref{tab:parameters} and show the parameters of each simulation underlying the map interpolation shown in the figure. Contours are shown in unit intevals, the dashed lines marking the $\Delta B^\ast=0$ iso-line. }
    \label{fig:du}
\end{figure}
The GQ model relies on the hypothesis that, provided $\Rey$ is high enough for the logarithmic law \eqref{eq:loglaw} to describe well the mean velocity profile, the quantity $\Delta B^*$ is a function of the control parameters only, and thus independent of the Reynolds number. 
This hypothesis is here tested using the $\Delta B^*$ maps for the DNS set at $\Rey_{\tau_0}=1000$ and $2000$. 
The maps are generated by applying the GQ model with the corresponding values of $\cfa$, $\cfr$ and $\DR$. 
The results, reported in figure~\ref{fig:du}, show maximum change of $\Delta B^*$ across $\Rey$ of only 0.36, with standard deviation 0.10. These values can be considered quite small, given that the maximum statistical uncertainty on the change of $\Delta B^*$ at 95\% confidence level is 0.24 across the map of figure \ref{fig:du}, and the mean absolute value is 0.17. 
This result thus confirms that the drag reduction effect barely changes with $\Rey$, once it is expressed in terms of $\Delta B^*$.
  
This additionally indicates that $\Rey_{\tau_0}=1000$ is sufficient to obtain a reasonably $\Rey$-independent estimate of $\Delta B^*$. This observation is also supported by the good agreement between the GQ16 data at $\Rey_{\tau_0} = 1000$ and the results by \cite{rouhi-etal-2023} obtained up to $\Rey_{\tau_0}=4000$ in relatively small domains.

\subsection{Monotonicity of $\DR$ with $\Rey$}
\label{sec:res_DR_Re}

\begin{figure}
\centering
\includegraphics[]{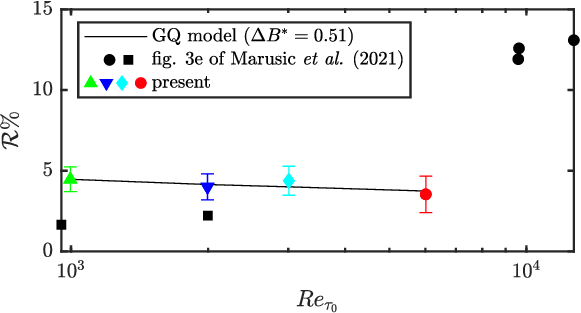}
	\caption{Drag reduction rate ($\DR$) as function of the reference friction Reynolds number ($\Rey_{\tau_0}$) for backward-travelling wave with parameters $A^+=5$, $\kappa^+=0.00078$ and $\omega^+=-0.0105$, close to the conditions considered by \cite{marusic-etal-2021}, i.e. $A^+ \approx 5$, $\kappa^+ \approx 0.0008$ and $\omega^+ \approx -0.0105$ (in their laboratory experiment the viscous-scaled parameters vary slightly with $\Rey$). The present results are denoted with coloured symbols (see table \ref{tab:parameters}); experimental data by \cite{marusic-etal-2021} are black solid circles, while open circles denote their LES numerical data; the straight line is the prediction of the GQ model \eqref{eq:gqmodel} corresponding to $\Delta B^*=0.51$ and to the values of $\cfr$ obtained from the uncontrolled simulations at the respective value of $\Rey_{\tau_0}$. The error bars have been determined as described in \S\ref{sec:methods}, corresponding to a 95\% confidence level.}
\label{fig:dr_re}
\end{figure}

The GQ model predicts that $\DR$ decreases monotonically with $\Rey$, however more slowly than the power-law decrease assumed in early studies~\citep{choi-xu-sung-2002, quadrio-ricco-2004, touber-leschziner-2012}. The decrease rate is less at higher $\Rey$ and for smaller $\DR$. Ample numerical and experimental evidence so far, including the results of the present study, support the predictions of the GQ model. 

Contrasting evidence that $\DR$ may instead increase with $\Rey$ has been recently provided from the combined laboratory and numerical efforts of \cite{marusic-etal-2021}. As shown in figure 3{\it e} of their paper, they found that $\DR$ obtained by backward-travelling waves at small values of $\kappa^+$ and $\omega^+$, namely $\kappa^+=0.0008$ and $\omega^+=-0.0105$, increases from 1.6\% at $\Rey_{\tau_0}\approx 1000$, as measured numerically in large-eddy simulation (LES) of open channel flow, up to 13.1\% at $\Rey_{\tau_0}\approx 12800$, as measured experimentally in a turbulent boundary layer. Since the actuator employed in their experiments yields a wave with a frequency-dependent amplitude and constant wavelength in physical units (30 cm), those authors could not exactly maintain the same value of viscous-scaled control parameters across the considered Reynolds number range. Specifically, the amplitude increased from $A^+=4.6$ at $\Rey_\tau = 9000$ to $A^+ = 5.7$ at $\Rey_\tau = 12800$ \cite[see table 1 in][]{chandran-etal-2023}. Furthermore, although the original figure 3{\it e} of \cite{marusic-etal-2021} reports a constant value of $\kappa^+=0.0008$ at all $\Rey$, we cannot reconcile it with the actuator wavelength being fixed in physical units for the experimental points. 

In the present work, we verify this contrasting evidence by studying the $\Rey$-dependence of $\DR$ across the largest range of Reynolds number tested so far via DNS. For this purpose, we consider StTW actuation at $\Rey_{\tau_0}=1000$, 2000, 3000 and 6000, with control parameters selected to match as closely as possible those reported in figure 3{\it e} of \cite{marusic-etal-2021}, namely $\kappa^+=0.00078$ and $\omega^+=-0.0104$. The wave amplitude is set to $A^+=5$, midway between the range of variation in their experiments.
Figure~\ref{fig:dr_re} compares our numerical results with the numerical and experimental results of \cite{marusic-etal-2021}. Our measurements still fit very well the prediction of the GQ model, and confirm an overall decreasing trend of $\DR$ with $\Rey$.

To verify whether the differences observed in figure~\ref{fig:dr_re} are due to the different Reynolds number range considered here and by \cite{marusic-etal-2021}, we advocate the work of \cite{chandran-etal-2023}. Those authors extended the experimental database of \cite{marusic-etal-2021} with additional data points, some of which at $\Rey_{\tau_0} \approx 6000$, i.e. the highest Reynolds number considered in the present study. Hence, we have precisely reproduced their actuated flow case with $\left\{A^+, \omega^+, \kappa^+, Re_{\tau_0} \right\} = \left\{2.5, -0.009, 0.0014, 6000 \right\}$, the remaining differences being the flow configuration (open channel vs. boundary layer), as well as actuation details (ideal harmonic actuation in numerical simulation vs. spatially discretised wave in experiment). 
This case also falls within the range of potential use for outer-scaled actuation according to \cite{deshpande-etal-2023c}, due to the comparatively large actuation period $T^+ = -700$ and wavelength $\lambda^+ \approx 4500$, similar to the case presented in figure \ref{fig:dr_re}. A drag reduction of $\DR = 2.3\% \pm 1.1\%$ is measured here, to be compared with $\DR = 6\%$ measured experimentally by \cite{chandran-etal-2023}. This finding hints at systematic differences between the present numerical simulations and the laboratory experiments of \cite{marusic-etal-2021} and \cite{chandran-etal-2023}. We reiterate that this is possibly due to irreducible differences in the flow and wall actuation setups, or even to the extreme challenges posed by laboratory experiments targeting such complex drag reduction strategies. We will go back to this important issue in \S\ref{sec:conclusion}. 
For the moment, the present data corroborate the expectation that $\DR$ decreases with $\Rey$ at the rate predicted by the GQ model.

\subsection{Net power savings at large values of $\Rey$}
\label{sec:res_ndr}

Net power saving $\NDR$ derives from the (positive or negative) balance between the power saved through drag-reducing control and the power required for wall actuation, hence
\begin{equation}
    \NDR = \DR - \frac{\pin}{\ppnot} \, ,
    \label{eq:ndr}
\end{equation}
where $\ppnot$ is the pumping power per unit wetted area in the uncontrolled case, which for constant $U_b$ reads  
\begin{equation}
    \ppnot = U_b \tau_{x_0} \, ,
    \label{eq:p0}
\end{equation}
and $\pin$ is the control input power per unit wetted area, expressed as:
\begin{equation}
    \pin = \aver{ w_w \tau_z }
         = \rho \nu  \left. \aver{w \frac{\partial w}{\partial y}}\right|_w 
         = \frac{\rho \nu}{2} \left. \frac{\dd}{\dd y} \aver{w w} \right|_{w}\, ,
    \label{eq:pin}
\end{equation}
where $\tau_z = \rho \nu (\partial w / \partial y)_w$ is the spanwise wall shear stress. 

Similarly to what done for $\DR$, the Reynolds-number dependence of $\NDR$ can also be predicted theoretically. Whereas $\DR$ is accurately expressed by the GQ model, the $\Rey$-dependence of $\pin / \ppnot$ can be easily expressed following \cite{ricco-quadrio-2008}, who noticed that this ratio is equivalent to $\pin^+ / \ppnot^+$. Since $\pin^+$ is very well approximated by the power $P_\ell^+$ required to generate the laminar transverse Stokes layer~\citep{quadrio-ricco-2011, gatti-quadrio-2013} --- which does not depend on $\Rey$ if the viscous-scaled parameters are kept constant --- the $\Rey$-dependence of $\pin / \ppnot$ comes only from $\ppnot^+ = U_b^+ = \sqrt{2/\cfr}$. By using the expression of $P_{\ell}^+$ by \cite{gatti-quadrio-2013}, we thus obtain
\begin{equation}
    \frac{\pin}{\ppnot} \approx \frac{P_{\ell}^+}{U_b^+} = \frac{(A^+)^2 (\kappa^+)^{1/3}}{2 U_b^+} \mathrm{Re} \left[ \mathrm{e}^{\pi i / 6} \frac{\mathrm{Ai}^\prime (\theta)}{\mathrm{Ai}(\theta)} \right] \, ,
    \label{eq:pin_ppnot_model}
\end{equation}
where $i$ is the imaginary unit, Re indicates the real part of a complex number, $\mathrm{Ai}$ is the Airy function of the first kind, $\mathrm{Ai}^\prime$ its derivative and $\theta = - \mathrm{e}^{\pi i / 6} (\kappa^+)^{1/3} \left( \omega^+/\kappa^+ + i\kappa^+ \right)$. Equation \eqref{eq:pin_ppnot_model} shows that $\pin^+ = U_b^+ \pin / \ppnot \approx P_\ell^+$, is a Reynolds-independent quantity for StTW parameters sufficiently far from the region of drag increase, where the approximation $\pin^+ \approx P_{\ell}^+$ is known to fail. As a result, it is sufficient to measure $\pin^+$ at a given Reynolds number, or estimate it via $P_{\ell}^+$, in order to retrieve $\pin / \ppnot$ at any Reynolds number, i.e. at any arbitrary $U_b^+=\sqrt{2/\cfr}$. Equation~\eqref{eq:pin_ppnot_model} shows that $\pin / \ppnot$ decreases with $\Rey$ as $1/U_b^+$, so that $\NDR$ can in fact increase with $\Rey$, provided the normalised actuation power decays with $\Rey$ faster than $\DR$.

\begin{figure}
    \includegraphics[]{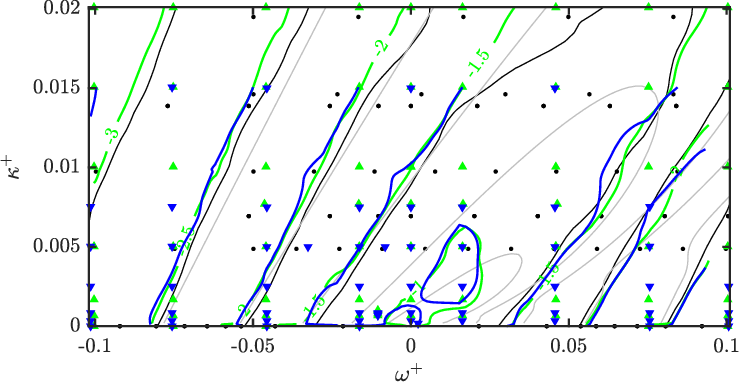}
	\caption{Maps of actuation power ($\pin^+$) as a function of the actuation parameters ($\omega^+$, $\kappa^+$), at $\Rey_{\tau_0}=1000$ \mbox{(
        \protect\tikz{\protect\draw[thick,re1000] (0pt,2pt) -- (20pt,2pt); 
                      \protect\draw[thick,white] (0pt,0pt) -- (20pt,0pt);}
        )} and $\Rey_{\tau_0}=2000$ \mbox{(
        \protect\tikz{\protect\draw[thick,re2000] (0pt,2pt) -- (20pt,2pt); 
                      \protect\draw[thick,white] (0pt,0pt) -- (20pt,0pt);}  
        )}. The symbols are colored after table \ref{tab:parameters} and show the parameters of each simulation underlying the map interpolation shown in the figure. Data by GQ16 
        \mbox{(
        \protect\tikz{\protect\draw[black] (0pt,2pt) -- (20pt,2pt); 
                      \protect\draw[white] (0pt,0pt) -- (20pt,0pt);} and black dots)}, and $\pin^+$ from equation \eqref{eq:pin_ppnot_model} \mbox{(
            \protect\tikz{\protect\draw[black!25] (0pt,2pt) -- (20pt,2pt); 
                          \protect\draw[white] (0pt,0pt) -- (20pt,0pt);}  
            )}  are also reported.
        }
    \label{fig:Pinplus}
\end{figure}
Figure \ref{fig:Pinplus} confirms that $\pin^+$ is indeed constant with $\Rey$ throughout the  investigated parameter space, included the drag-increasing regime, where $\pin^+$ and $P_\ell^+$ do differ and the former can only be measured empirically. The GQ16 dataset well aligns with the present data, the lacking information for $0 < \kappa^+ \leq 0.005$ notwithstanding. 

\begin{figure}
    \includegraphics[]{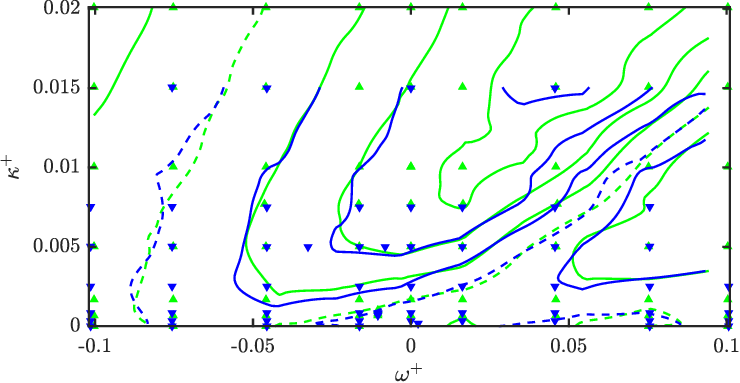}
	\caption{Maps of net power saving ($\NDR$) as a function of the actuation parameters ($\omega^+$, $\kappa^+$), at $\Rey_{\tau_0}=1000$ \mbox{(
        \protect\tikz{\protect\draw[thick,re1000] (0pt,2pt) -- (20pt,2pt); 
                      \protect\draw[thick,white] (0pt,0pt) -- (20pt,0pt);}
        )} and $\Rey_{\tau_0}=2000$ \mbox{(
        \protect\tikz{\protect\draw[thick,re2000] (0pt,2pt) -- (20pt,2pt); 
                      \protect\draw[thick,white] (0pt,0pt) -- (20pt,0pt);}  
        )}. The symbols are colored after table \ref{tab:parameters} and show the parameters of each simulation underlying the map interpolation shown in the figure. Contour lines are shown in intervals of 5\%, the dashed lines denoting the $\NDR=0$ iso-line.
        }
    \label{fig:NDR}
\end{figure}
The net power saving at $\Rey_{\tau_0}=1000$ and $2000$ is reported in figure~\ref{fig:NDR}. Overall, the contours of $\NDR$ do not change significantly, since degradation of $\DR$ is compensated by reduction of the actuation input power. Larger differences are observed for nearly optimal $\NDR$ (see the $\NDR=15\%$ iso-line in figure~\ref{fig:NDR}), in a region which shrinks and shifts towards higher $\kappa^+$ at higher $\Rey$. This can be explained by the stronger decay of $\DR$ in this region (as predicted by the GQ model due to larger $\DR$) and by the comparatively small value of $\pin / \ppnot$, which causes $\NDR$ to 
have similar $\Rey$-dependence as $\DR$. 
 
GQ16 noticed that at $\Rey_{\tau_0} \approx 1000$ and $A^+=5.5$ the locus of near-optimum new power saving ($\NDR = 15\%$) extends along the ridge of maximum $\DR$ between $\kappa^+ = 0.0085$ and $0.04$, the maximum being at $\left\{\omega^+, \kappa^+\right\} = \left\{0.093, 0.026 \right\}$. This implies that the point of maximum $\NDR$ might reside outside of the parameter space considered in figure~\ref{fig:NDR}, for both Reynolds numbers under scrutiny here. 

\begin{figure}
\centering
\includegraphics[]{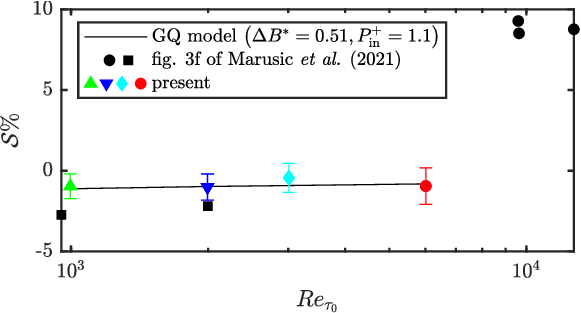}
	\caption{Net power saving ($\NDR$) as function of reference friction Reynolds number ($\Rey_{\tau_0}$) for backward-travelling waves with the same parameters considered by \cite{marusic-etal-2021}. The present data are indicated with colored symbols (see table \ref{tab:parameters}); data by \cite{marusic-etal-2021} are black solid circles; the straight line is the theoretical prediction obtained by combining the GQ model \eqref{eq:gqmodel} for $\Delta B^\ast=0.51$ with equation~\eqref{eq:pin_ppnot_model} for $\pin^+=1.1$ and the values of $\cfr$ obtained from the uncontrolled simulations at the respective value of $\Rey_{\tau_0}$.}
\label{fig:s_re}
\end{figure}

As done for the drag reduction in figure \ref{fig:dr_re}, the variation of $\NDR$ with $\Rey$ is shown in figure \ref{fig:s_re}, for the same parameters considered by \cite{marusic-etal-2021}. Interestingly, $\NDR$ is observed to increase with $\Rey$ at this combination of parameters, essentially due to the shrinking of the negative $\pin / \ppnot$ contribution and to the relatively constant $\DR$. The increase of $\NDR$ is compatible with the theoretical prediction that can be obtained by combining the GQ model of equation~\eqref{eq:gqmodel} with the prediction for $\pin / \ppnot$ of equation \eqref{eq:pin_ppnot_model}. The differences between the present numerical database and the laboratory experiments of \cite{marusic-etal-2021}, previously noted for $\DR$, are confirmed here. 

The present results enable a better understanding of the available literature data. For instance, by comparing the numerical data by \cite{rouhi-etal-2023}, which consider StTW at small wavelengths (due to the restricted domain size) and relatively large amplitude $A^+ = 12$ and frequencies, with their experimental data, which consider backward-travelling waves at larger wavelengths but smaller amplitudes of $A^+ \approx 5$ and frequencies, \cite{chandran-etal-2023} conclude that mostly low-frequency forcing $\left| \omega^+ \right| <0.018$ is capable to achieve positive $\NDR$, despite the moderate values of $\DR$. This conclusion is observed here to be an artifact of the comparison between StTW at different amplitudes: according to  GQ16 it is known that already at $Re_{\tau_0}=1000$ no positive $\NDR$ can be achieved via StTW for amplitudes $A^+ \gtrapprox 14$. The present data clearly show that the observation of GQ16 is valid also if smaller values of wavenumbers and frequencies are considered: the locus of maximum $\NDR$ in the $\left\{\omega, \kappa \right\}$-space essentially coincides with the one of maximum $\DR$, and it shifts towards larger $\left\{\omega, \kappa \right\}$ for increasing values of  $\Rey$ rather than to smaller ones, if the comparison among various $\Rey$ is performed at a constant value of $A^+$ close to the optimal $A^+ \approx 6$ identified by GQ16.  \section{Concluding discussion}
\label{sec:conclusion}

In the present work we have addressed the Reynolds-number dependence of skin-friction drag reduction induced by spanwise forcing, in terms of both drag reduction rate $\DR$ and net power saving $\NDR$. In particular, we have focused on streamwise-travelling waves of spanwise wall velocity \citep[StTW, ][]{quadrio-ricco-viotti-2009}. 
A new database of high-fidelity direct numerical simulations (DNS) of turbulent open channel flow with and without StTW has been generated for $\Rey_{\tau_0}=1000$, 2000, 3000 and 6000. This is the widest Reynolds-number range considered so far in numerical experiments with spanwise forcing, and reduces the gap from the highest value of $\Rey_{\tau_0}$ considered in analogous laboratory experiments~\citep{chandran-etal-2023} to a factor of 2.5. 

The main outcome of the present study is to confirm the validity of the predictive model for drag reduction proposed by \cite{gatti-quadrio-2016} and its underlying hypothesis. 
The present data corroborate the observation that the parameter $\Delta B^*$, which quantifies the control-induced velocity shift in actual viscous units ``$*$'' at matched $y^*$ with respect to the non-actuated flow, is a $\Rey$-independent measure of drag reduction when the Reynolds number is sufficiently large for the logarithmic law to apply.
We have shown that $\Rey_{\tau_0} \gtrapprox 1000$ is sufficient for $\Delta B^*$ to become nearly $\Rey$-independent, since no statistically significant differences have been measured between the $\Rey_{\tau_0}=1000$ and $\Rey_{\tau_0}=2000$ cases, for a wide range of actuation parameters, and up to $\Rey_{\tau_0}=6000$ for one selected combination of actuation parameters. 

This key result implies that drag reduction induced by StTW at a given combination of $\left\{A^+, \omega^+, \kappa^+ \right\}$ is bound to monotonically decrease with the Reynolds number, at a rate that depends on $\DR$ itself and on (the inverse square root of) the skin-friction coefficient $\cfr$ of the uncontrolled flow, as embodied in the GQ model; see equation \eqref{eq:gqmodel}. Fortunately, 
the decay rate is less severe than the power law $\DR \sim Re_{\tau_0}^{-0.2}$ suggested empirically in early studies on spanwise wall oscillations~\citep{choi-xu-sung-2002,touber-leschziner-2012}, conveying that significant drag reduction can still be achieved at very high $\Rey$. 

The increase of drag reduction with the Reynolds number observed by \cite{marusic-etal-2021} with actuation 
parameters corresponding to the outer-scaled actuation is not confirmed by our numerical experiments with $\left\{A^+ = 5, \omega^+ = -0.0104, \kappa^+= 0.00078 \right\}$ in turbulent open channels. On the contrary, the present results follow well the prediction of the GQ model, and show a very mild decrease of $\DR$ with $\Rey$ for these specific parameters. 
While the observation of $\DR$ increasing with $\Rey$ is indeed surprising and unique in literature, we can only speculate on the reasons behind this discrepancy.

On the one hand, the difference in the flow setup considered here and in \cite{marusic-etal-2021} (open channel vs. boundary layer) could affect the Reynolds-number dependence of $\DR$. In this respect, \cite{skote-2014} applied StTW to numerical turbulent boundary layers at low $\Rey$ and noted that the K\'arm\'an constant can increase in the presence of drag-reduction effects. This could affect the $\Rey$-dependency of $\DR$, since the GQ model assumes constancy of $k$. The experimental data of \cite{chandran-etal-2023}, however, do not support such an effect.
On the other hand, \cite{marusic-etal-2021} and later \cite{chandran-etal-2023} implemented a spatially discrete form of the StTW, similarly to \cite{auteri-etal-2010}, and synthesised harmonic waves by independently moving stripes with finite width. \cite{auteri-etal-2010} and, more recently, \cite{gallorini-quadrio-2024} addressed the effects of the wave discretisation on the achievable drag reduction. Owing to discretization, the turbulent flow perceives a number of higher Fourier harmonics of the discrete piecewise-constant wave, as if multiple waves with different parameters were applied. 
As a result, quantitative comparison between the ideally continuous and piecewise-constant forcing is not trivial, and some discrete waves far from the optimal forcing parameters can outperform the corresponding ideal sinusoidal waveform, whenever part of the harmonic content of the discrete wave falls in high-$\DR$ regions of the drag reduction map. 
Finally, the conclusion of \cite{marusic-etal-2021} that $\DR$ increases with $\Rey$ hinges on comparison of data obtained with different methods. In particular, the low-$\Rey$ data were obtained from LES of turbulent open channel flow in relatively small domains with continuous StTW applied at the wall, whereas the high-$\Rey$ data were obtained from boundary layer experiments with discrete StTW. Differences in numerical and experimental uncertainties can further complicate the comparison.
Whereas the above speculations remain to be verified in future studies, the present results support the claim that ideal StTW applied in turbulent open channels are neither expected nor observed to yield an increase of drag reduction with increasing $Re$, for any combination of wave parameters that are kept constant in viscous units. 

\begin{figure}
\centering
\includegraphics[]{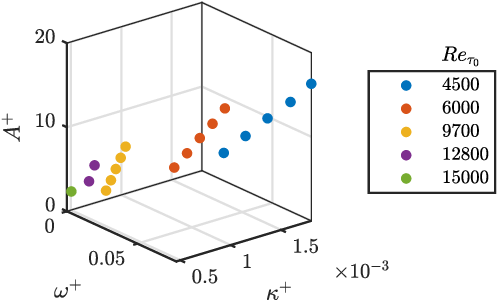}
	\caption{Wavenumber ($\kappa^+$), angular frequency ($\omega^+$) and amplitude ($A^+$) for StTW actuation considered by \cite{chandran-etal-2023}, for different values of $\Rey_{\tau_0}$. \label{fig:chandran}}
\end{figure}

Lastly, we also confirm that the Reynolds-number dependence of the net power saving $\NDR = \DR - \pin / \ppnot$ is in line with theoretical predictions. Whereas $\DR$ directly derives from the GQ model, $\pin / \ppnot$ can be obtained directly from $\pin^+ = U_b^+ \pin / \ppnot$, which is known to be $\Rey$-independent~\citep{gatti-quadrio-2013}.
Interestingly, we have found that $\pin^+$ does not change with $Re$ throughout the drag-reduction map, not only in those regions where $\pin^+$ is known to be well approximated by $P_\ell^+$, i.e. the value obtained from the laminar generalised Stokes layer solution. In other words, the ideal viscous scaling of $\pin^+$ is retained even close to the valley of drag increase, where turbulence is known to interact with the generalised Stokes layer generated by StTW actuation. 
This result, as already discussed in \cite{gatti-quadrio-2013, gatti-quadrio-2016}, has two main implications. Firstly, in the portion of the StTW parameter space where $\NDR$ is maximum, $\NDR$ is dominated by $\DR$ and hence exhibits similar $\Rey$-dependence; here $\NDR$ decreases with $\Rey$ at a rate which is slightly less than $\DR$. Secondly, for StTW parameters far from the optimum, both $\DR$ and $\pin/ \ppnot$ contribute to $\NDR$. In this case, the normalised control cost may decrease with $\Rey$ at a faster rate than $\DR$, so that $\NDR$ can actually increase with $\Rey$. However, this can occur only in regions of non-optimal values of $\NDR$. 
Hence, we argue that the observation by \cite{chandran-etal-2023} that only low-frequency, low-wavenumber forcing can achieve positive $\NDR$ at high $\Rey$ may be an artifact due to the properties of their experimental setup, in which the same region of the viscous-scaled parameter space cannot be spanned for different values of $\Rey$ (see figure~\ref{fig:chandran}). Indeed, those authors can only achieve the optimal values of $A^+$ at the highest values of $\Rey$, at which only low $\omega^+$ and $\kappa^+$ are possible owing to the small space- and time-scales of the turbulent flow. The more systematic scan of the StTW parameter space carried out in the present study shows that the loci of optimal $\NDR$ and $\DR$ roughly coincide in the $\left\{\omega^+, \kappa^+ \right\}$ plane. 

\section*{Acknowledgments}
This work was supported by the EuroHPC Joint Undertaking (JU) under grant EHPC-EXT-2022E01-054 on the Leonardo Booster machine at CINECA.

\section*{Funding} 
This research received no specific grant from any funding agency, commercial or not-for-profit sectors.

\section*{Declaration of Interests} 
The authors report no conflict of interest.

\appendix
\section{Dataset details}
\label{sec:appendix}

This Appendix reports the combination of the StTW control parameters of the simulations performed to produce the present dataset, together with the main quantities of interest. Tables \ref{tab:ret1000}, \ref{tab:ret2000}, \ref{tab:ret3000} and \ref{tab:ret6000} are for $Re_{\tau_0} = 1000$, $Re_{\tau_0} = 2000$, $Re_{\tau_0} = 3000$ and $Re_{\tau_0} = 6000$ respectively.
\clearpage{}\begin{center}
   \setlength{\tabcolsep}{7pt}
   \begin{longtable}{@{\extracolsep{\fill}}lcS@{}S@{}cS@{}cS@{}S@{}}
      \hline
      Case & {$A/U_b$} & {$A^+$} & {$\omega h / U_b$} & {$\omega^+ \times 10^2$} & {$\kappa h$} & {$\kappa^+ \times 10^2$} & $\DR \%$ & $\NDR \%$ \\
      \hline\\
1   &   0.248  &    5.0  &   -4.98  &               -10.03  &          0  &  0.000 &       12.2  &  -1.6\\
2   &   0.248  &    5.0  &   -4.98  &               -10.03  &        0.67  &  0.067 &       12.7  &  -1.2\\
3   &   0.248  &    5.0  &   -4.98  &               -10.03  &        1.7  &  0.167 &       12.8  &  -1.3\\
4   &   0.248  &    5.0  &   -4.98  &               -10.03  &          5  &  0.502 &       12.6  &  -1.9\\
5   &   0.248  &    5.0  &   -4.98  &               -10.03  &         10  &  1.004 &       11.4  &  -3.6\\
6   &   0.248  &    5.0  &   -4.98  &               -10.03  &         15  &  1.506 &       9.8  &  -5.8\\
7   &   0.248  &    5.0  &   3.73  &               7.53  &          0  &  0.000 &       13.3  &  1.4\\
8   &   0.248  &    5.0  &   3.73  &               7.53  &        0.67  &  0.067 &       12.5  &  0.8\\
9   &   0.248  &    5.0  &   3.73  &               7.53  &        1.7  &  0.167 &       10.5  &  -1.0\\
10   &   0.248  &    5.0  &   4.98  &               10.03  &          0  &  0.000 &       12.2  &  -1.6\\
11   &   0.248  &    5.0  &   3.73  &               7.53  &          5  &  0.502 &       -1.0  &  -11.5\\
12   &   0.248  &    5.0  &   3.73  &               7.53  &         10  &  1.004 &       6.7  &  -2.0\\
13   &   0.248  &    5.0  &   3.73  &               7.53  &         15  &  1.506 &       19.3  &  12.9\\
14   &   0.248  &    5.0  &   4.98  &               10.03  &        0.67  &  0.067 &       12.0  &  -1.7\\
15   &   0.248  &    5.0  &   4.98  &               10.03  &        1.7  &  0.167 &       12.0  &  -1.5\\
16   &   0.248  &    5.0  &   4.98  &               10.03  &          5  &  0.502 &       5.4  &  -7.5\\
17   &   0.248  &    5.0  &   4.98  &               10.03  &         10  &  1.004 &       -1.8  &  -13.8\\
18   &   0.248  &    5.0  &   4.98  &               10.03  &         15  &  1.506 &       10.0  &  0.1\\
19   &   0.248  &    5.0  &   -2.30  &               -4.64  &        7.7  &  0.770 &       16.3  &  5.3\\
20   &   0.248  &    5.0  &   -0.82  &               -1.65  &        7.7  &  0.770 &       19.8  &  11.4\\
21   &   0.248  &    5.0  &   -0.00  &               -0.00  &        7.7  &  0.770 &       21.6  &  14.8\\
22   &   0.248  &    5.0  &   0.82  &               1.65  &        7.7  &  0.770 &       20.9  &  15.7\\
23   &   0.248  &    5.0  &   2.30  &               4.64  &        7.7  &  0.770 &       10.3  &  4.9\\
24   &   0.248  &    5.0  &   -2.28  &               -4.59  &          0  &  0.000 &       10.2  &  1.2\\
25   &   0.248  &    5.0  &   -0.81  &               -1.63  &          0  &  0.000 &       2.8  &  -3.1\\
26   &   0.248  &    5.0  &   -2.28  &               -4.59  &        0.67  &  0.067 &       11.8  &  2.5\\
27   &   0.248  &    5.0  &   -0.81  &               -1.63  &        0.67  &  0.067 &       5.3  &  -0.7\\
28   &   0.248  &    5.0  &   -0.00  &               -0.00  &        0.67  &  0.067 &       1.0  &  -3.4\\
29   &   0.248  &    5.0  &   0.81  &               1.63  &        0.67  &  0.067 &       0.8  &  -4.8\\
30   &   0.248  &    5.0  &   2.28  &               4.59  &        0.67  &  0.067 &       8.0  &  -0.8\\
31   &   0.248  &    5.0  &   -2.28  &               -4.59  &        1.7  &  0.167 &       14.3  &  4.8\\
32   &   0.248  &    5.0  &   -0.81  &               -1.63  &        1.7  &  0.167 &       10.3  &  3.9\\
33   &   0.248  &    5.0  &   -0.00  &               -0.00  &        1.7  &  0.167 &       4.7  &  -0.1\\
34   &   0.248  &    5.0  &   0.81  &               1.63  &        1.7  &  0.167 &       2.2  &  -2.1\\
35   &   0.248  &    5.0  &   2.28  &               4.59  &        1.7  &  0.167 &       3.5  &  -5.0\\
36   &   0.248  &    5.0  &   -2.28  &               -4.59  &          5  &  0.502 &       16.6  &  6.2\\
37   &   0.248  &    5.0  &   -0.81  &               -1.63  &          5  &  0.502 &       18.1  &  10.5\\
38   &   0.248  &    5.0  &   -0.00  &               -0.00  &          5  &  0.502 &       17.5  &  11.6\\
39   &   0.248  &    5.0  &   0.81  &               1.63  &          5  &  0.502 &       13.5  &  9.1\\
40   &   0.248  &    5.0  &   2.28  &               4.59  &          5  &  0.502 &       3.0  &  -3.8\\
41   &   0.248  &    5.0  &   2.28  &               4.59  &          5  &  0.502 &       3.0  &  -3.8\\
42   &   0.248  &    5.0  &   -2.28  &               -4.59  &         10  &  1.004 &       15.6  &  4.2\\
43   &   0.248  &    5.0  &   -0.81  &               -1.63  &         10  &  1.004 &       19.0  &  10.0\\
44   &   0.248  &    5.0  &   -0.00  &               -0.00  &         10  &  1.004 &       20.8  &  13.3\\
45   &   0.248  &    5.0  &   0.81  &               1.63  &         10  &  1.004 &       22.5  &  16.5\\
46   &   0.248  &    5.0  &   2.28  &               4.59  &         10  &  1.004 &       18.2  &  13.4\\
47   &   0.248  &    5.0  &   -2.28  &               -4.59  &         15  &  1.506 &       14.0  &  1.8\\
48   &   0.248  &    5.0  &   -0.81  &               -1.63  &         15  &  1.506 &       17.4  &  7.4\\
49   &   0.248  &    5.0  &   -0.00  &               -0.00  &         15  &  1.506 &       18.9  &  10.2\\
50   &   0.248  &    5.0  &   0.81  &               1.63  &         15  &  1.506 &       20.5  &  13.0\\
51   &   0.248  &    5.0  &   2.28  &               4.59  &         15  &  1.506 &       24.3  &  18.8\\
52   &   0.228  &    4.6  &   -0.00  &               -0.00  &         10  &  1.004 &       19.7  &  13.3\\
53   &   0.248  &    5.0  &   -0.52  &               -1.05  &        0.67  &  0.067 &       3.9  &  -1.0\\
54   &   0.248  &    5.0  &   -0.52  &               -1.05  &          1  &  0.100 &       5.4  &  -0.2\\
55   &   0.248  &    5.0  &   -3.73  &               -7.53  &          0  &  0.000 &       12.3  &  0.4\\
56   &   0.248  &    5.0  &   -3.73  &               -7.53  &        0.67  &  0.067 &       13.4  &  1.3\\
57   &   0.248  &    5.0  &   -3.73  &               -7.53  &        1.7  &  0.167 &       13.2  &  1.0\\
58   &   0.248  &    5.0  &   -3.73  &               -7.53  &          5  &  0.502 &       13.8  &  1.0\\
59   &   0.248  &    5.0  &   -3.73  &               -7.53  &         10  &  1.004 &       12.3  &  -1.2\\
60   &   0.248  &    5.0  &   -3.73  &               -7.53  &         15  &  1.506 &       11.8  &  -2.4\\
61   &   0.248  &    5.0  &   -4.98  &               -10.04  &         20  &  2.008 &       8.5  &  -7.7\\
62   &   0.248  &    5.0  &   -3.73  &               -7.53  &         20  &  2.008 &       10.8  &  -4.0\\
63   &   0.248  &    5.0  &   -2.28  &               -4.59  &         20  &  2.008 &       13.0  &  0.1\\
64   &   0.248  &    5.0  &   -0.81  &               -1.63  &         20  &  2.008 &       15.6  &  4.7\\
65   &   0.248  &    5.0  &   -0.00  &               -0.00  &         20  &  2.008 &       17.3  &  7.5\\
66   &   0.248  &    5.0  &   0.81  &               1.63  &         20  &  2.008 &       19.5  &  10.9\\
67   &   0.248  &    5.0  &   2.28  &               4.59  &         20  &  2.008 &       22.9  &  16.2\\
68   &   0.248  &    5.0  &   3.73  &               7.53  &         20  &  2.008 &       23.8  &  17.7\\
69   &   0.248  &    5.0  &   4.98  &               10.04  &         20  &  2.008 &       19.8  &  11.8\\
70   &   0.248  &    5.0  &   -0.52  &               -1.05  &        0.78  &  0.078 &       4.5  &  -1.0\\
      \hline
      \caption{List of the controlled simulations carried out at $Re_{\tau_0} =1000 $.} \label{tab:ret1000} \\
   \end{longtable}
\end{center}
\clearpage{}
\clearpage{}\begin{center}
   \setlength{\tabcolsep}{7pt}
   \begin{longtable}{@{\extracolsep{\fill}}lcS@{}S@{}cS@{}cS@{}S@{}}
      \hline
      Case & {$A/U_b$} & {$A^+$} & {$\omega h / U_b$} & {$\omega^+ \times 10^2$} & {$\kappa h$} & {$\kappa^+ \times 10^2$} & $\DR \%$ & $\NDR \%$ \\
      \hline\\
1   &   0.3  &    6.5  &   0.21  &               0.23  &        0.33  &  0.017 &       2.7  &  -2.1\\
2   &   0.228  &    5.0  &   -1.50  &               -1.63  &          0  &  0.000 &       1.9  &  -3.5\\
3   &   0.228  &    5.0  &   -4.19  &               -4.56  &          0  &  0.000 &       9.5  &  1.3\\
4   &   0.228  &    5.0  &   1.50  &               1.63  &          5  &  0.250 &       3.6  &  -0.4\\
5   &   0.228  &    5.0  &   -1.50  &               -1.63  &        0.67  &  0.033 &       5.2  &  -0.2\\
6   &   0.228  &    5.0  &   -4.19  &               -4.56  &        0.67  &  0.033 &       10.6  &  2.3\\
7   &   0.228  &    5.0  &   -0.00  &               -0.00  &        0.67  &  0.033 &       3.6  &  -0.1\\
8   &   0.228  &    5.0  &   4.19  &               4.56  &        0.67  &  0.033 &       9.0  &  0.8\\
9   &   0.228  &    5.0  &   1.50  &               1.63  &        0.67  &  0.033 &       3.2  &  -2.0\\
10   &   0.228  &    5.0  &   -1.50  &               -1.63  &          5  &  0.250 &       12.8  &  6.6\\
11   &   0.228  &    5.0  &   -4.19  &               -4.56  &        1.7  &  0.083 &       13.0  &  4.5\\
12   &   0.228  &    5.0  &   -0.00  &               -0.00  &        1.7  &  0.083 &       1.9  &  -2.2\\
13   &   0.228  &    5.0  &   4.19  &               4.56  &        1.7  &  0.083 &       8.3  &  0.3\\
14   &   0.228  &    5.0  &   1.50  &               1.63  &        1.7  &  0.083 &       -0.2  &  -5.2\\
15   &   0.228  &    5.0  &   -1.50  &               -1.63  &         10  &  0.500 &       18.2  &  11.3\\
16   &   0.228  &    5.0  &   -4.19  &               -4.56  &          5  &  0.250 &       16.0  &  7.1\\
17   &   0.228  &    5.0  &   -0.00  &               -0.00  &          5  &  0.250 &       9.5  &  4.9\\
18   &   0.228  &    5.0  &   4.19  &               4.56  &          5  &  0.250 &       0.7  &  -6.9\\
19   &   0.228  &    5.0  &   -1.50  &               -1.63  &        1.7  &  0.083 &       6.7  &  1.1\\
20   &   0.228  &    5.0  &   -4.19  &               -4.56  &         10  &  0.500 &       16.0  &  6.6\\
21   &   0.228  &    5.0  &   4.19  &               4.56  &         10  &  0.500 &       3.2  &  -3.1\\
22   &   0.228  &    5.0  &   -1.50  &               -1.63  &         15  &  0.749 &       18.0  &  10.4\\
23   &   0.228  &    5.0  &   -4.19  &               -4.56  &         15  &  0.749 &       15.9  &  5.9\\
24   &   0.228  &    5.0  &   -0.00  &               -0.00  &         15  &  0.749 &       19.7  &  13.5\\
25   &   0.228  &    5.0  &   4.19  &               4.56  &         15  &  0.749 &       11.5  &  6.5\\
26   &   0.228  &    5.0  &   1.50  &               1.63  &         15  &  0.749 &       19.3  &  14.6\\
27   &   0.228  &    5.0  &   -0.75  &               -0.81  &         10  &  0.500 &       17.4  &  11.2\\
28   &   0.228  &    5.0  &   -0.00  &               -0.00  &         30  &  1.499 &       18.5  &  10.5\\
29   &   0.228  &    5.0  &   -2.99  &               -3.26  &         10  &  0.500 &       17.3  &  8.9\\
30   &   0.228  &    5.0  &   1.50  &               1.63  &         10  &  0.500 &       12.8  &  8.7\\
31   &   0.228  &    5.0  &   -0.00  &               -0.00  &         10  &  0.500 &       16.5  &  11.2\\
32   &   0.228  &    5.0  &   -0.95  &               -1.05  &        1.7  &  0.084 &       4.3  &  -0.7\\
33   &   0.228  &    5.0  &   -0.95  &               -1.05  &        1.3  &  0.067 &       3.1  &  -1.9\\
34   &   0.228  &    5.0  &   -4.19  &               -4.56  &         30  &  1.499 &       13.0  &  1.8\\
35   &   0.228  &    5.0  &   4.19  &               4.56  &         30  &  1.499 &       22.1  &  17.0\\
36   &   0.228  &    5.0  &   9.24  &               10.06  &          0  &  0.000 &       10.9  &  -1.9\\
37   &   0.228  &    5.0  &   9.24  &               10.06  &        0.67  &  0.033 &       11.4  &  -1.3\\
38   &   0.228  &    5.0  &   9.24  &               10.06  &        1.7  &  0.083 &       11.2  &  -1.4\\
39   &   0.228  &    5.0  &   9.24  &               10.06  &          5  &  0.250 &       10.0  &  -2.4\\
40   &   0.228  &    5.0  &   9.24  &               10.14  &         10  &  0.501 &       4.0  &  -7.9\\
41   &   0.228  &    5.0  &   9.24  &               10.14  &         15  &  0.752 &       -4.0  &  -15.2\\
42   &   0.228  &    5.0  &   -9.24  &               -10.14  &          0  &  0.000 &       10.7  &  -2.3\\
43   &   0.228  &    5.0  &   -9.24  &               -10.14  &        0.67  &  0.033 &       10.6  &  -2.1\\
44   &   0.228  &    5.0  &   -9.24  &               -10.14  &        1.7  &  0.084 &       11.1  &  -1.7\\
45   &   0.228  &    5.0  &   -9.24  &               -10.14  &          5  &  0.251 &       11.5  &  -1.5\\
46   &   0.228  &    5.0  &   -9.24  &               -10.14  &         10  &  0.501 &       10.5  &  -2.9\\
47   &   0.228  &    5.0  &   -9.24  &               -10.14  &         15  &  0.752 &       10.4  &  -3.2\\
48   &   0.228  &    5.0  &   -6.89  &               -7.56  &          0  &  0.000 &       11.9  &  0.9\\
49   &   0.228  &    5.0  &   -6.89  &               -7.56  &        0.67  &  0.033 &       11.6  &  0.6\\
50   &   0.228  &    5.0  &   -6.89  &               -7.56  &        1.7  &  0.084 &       12.0  &  1.0\\
51   &   0.228  &    5.0  &   -6.89  &               -7.56  &          5  &  0.251 &       13.1  &  1.7\\
52   &   0.228  &    5.0  &   -6.89  &               -7.56  &         10  &  0.501 &       12.4  &  0.7\\
53   &   0.228  &    5.0  &   -6.89  &               -7.56  &         15  &  0.752 &       12.5  &  0.4\\
54   &   0.228  &    5.0  &   6.89  &               7.56  &          0  &  0.000 &       12.5  &  1.6\\
55   &   0.228  &    5.0  &   6.89  &               7.56  &        0.67  &  0.033 &       11.3  &  0.4\\
56   &   0.228  &    5.0  &   6.89  &               7.56  &        1.7  &  0.084 &       10.9  &  0.1\\
57   &   0.228  &    5.0  &   6.89  &               7.56  &          5  &  0.251 &       7.0  &  -3.3\\
58   &   0.228  &    5.0  &   6.89  &               7.56  &         10  &  0.501 &       -2.1  &  -11.8\\
59   &   0.228  &    5.0  &   6.89  &               7.56  &         15  &  0.752 &       -0.2  &  -9.5\\
60   &   0.228  &    5.0  &   -6.89  &               -7.56  &         30  &  1.504 &       11.1  &  -1.9\\
61   &   0.228  &    5.0  &   -0.95  &               -1.05  &        1.6  &  0.078 &       4.0  &  -1.0\\
      \hline
      \caption{List of the controlled simulations carried out at $Re_{\tau_0} =2000 $.} \label{tab:ret2000} \\
   \end{longtable}
\end{center}
\clearpage{}
\clearpage{}\begin{center}
   \setlength{\tabcolsep}{7pt}
   \begin{longtable}{@{\extracolsep{\fill}}lcS@{}S@{}cS@{}cS@{}S@{}}
      \hline
      Case & {$A/U_b$} & {$A^+$} & {$\omega h / U_b$} & {$\omega^+ \times 10^2$} & {$\kappa h$} & {$\kappa^+ \times 10^2$} & $\DR \%$ & $\NDR \%$ \\
      \hline\\
1   &   0.219  &    5.0  &   -2.16  &               -1.65  &        0.67  &  0.022 &       3.1  &  -2.1\\
2   &   0.219  &    5.0  &   -0.00  &               -0.00  &        0.67  &  0.022 &       1.4  &  -2.1\\
3   &   0.219  &    5.0  &   2.16  &               1.65  &        0.67  &  0.022 &       1.3  &  -3.8\\
4   &   0.219  &    5.0  &   2.16  &               1.65  &        1.7  &  0.056 &       -0.5  &  -5.5\\
5   &   0.218  &    5.0  &   -1.37  &               -1.05  &        2.3  &  0.078 &       4.4  &  -0.4\\
6   &   0.218  &    5.0  &   -1.37  &               -1.05  &        2.7  &  0.089 &       4.7  &  -0.1\\
      \hline
      \caption{List of the controlled simulations carried out at $Re_{\tau_0} =3000 $.} \label{tab:ret3000} \\
   \end{longtable}
\end{center}
\clearpage{}
\clearpage{}\begin{center}
   \setlength{\tabcolsep}{7pt}
   \begin{longtable}{@{\extracolsep{\fill}}lcS@{}S@{}cS@{}cS@{}S@{}}
      \hline
      Case & {$A/U_b$} & {$A^+$} & {$\omega h / U_b$} & {$\omega^+ \times 10^2$} & {$\kappa h$} & {$\kappa^+ \times 10^2$} & $\DR \%$ & $\NDR \%$ \\
      \hline\\
1   &   0.203  &    5.0  &   -2.55  &               -1.04  &        4.7  &  0.078 &       3.5  &  -0.9\\
2   &   0.101  &    2.5  &   -2.19  &               -0.90  &        8.3  &  0.139 &       2.3  &  1.2\\
      \hline
      \caption{List of the controlled simulations carried out at $Re_{\tau_0} =6000 $.} \label{tab:ret6000} \\
   \end{longtable}
\end{center}
\clearpage{}

\end{document}